%% file: arxiv.tex
\documentclass[onecolumn,aps,pra,superscriptaddress,10pt,preprint]{revtex4-2}
\usepackage{graphicx}
\usepackage{dcolumn}
\usepackage{bm}
\usepackage{hyperref}
\usepackage{stmaryrd}
\usepackage{soul}
\hypersetup{colorlinks,allcolors=blue}

\newcommand{\Tc}{$T_\mathrm{c}$}
\newcommand{\Tg}{$T_\mathrm{G}$}
\newcommand{\Po}{$P_\mathrm{O_2}$}
\newcommand{\AO}{Al$_2$O$_3$}
\newcommand{\dC}{$^\circ$C}
\newcommand{\RH}{$R_\mathrm{H}$}
\newcommand{\rr}{$\rho_\mathrm{0}$}

\begin{document}

\title{Thermally-Activated Epitaxy of NbO}
\author{Sandra~Glotzer}
\affiliation{Department of Applied Physics and Materials Science, California Institute of Technology, Pasadena, California 91125, USA.}
\affiliation{Institute for Quantum Information and Matter, California Institute of Technology, Pasadena, California 91125, USA.}

\author{Jeong~Rae~Kim}
\affiliation{Department of Applied Physics and Materials Science, California Institute of Technology, Pasadena, California 91125, USA.}
\affiliation{Institute for Quantum Information and Matter, California Institute of Technology, Pasadena, California 91125, USA.}

\author{Joseph~Falson}
\email{falson@caltech.edu}
\affiliation{Department of Applied Physics and Materials Science, California Institute of Technology, Pasadena, California 91125, USA.}
\affiliation{Institute for Quantum Information and Matter, California Institute of Technology, Pasadena, California 91125, USA.}


\begin{abstract}


We demonstrate a thermally-activated epitaxy window for the growth of NbO at temperatures exceeding 1000 \dC. NbO films grown in this mode display superior structural and transport properties, which are reproducible across a window of oxygen partial pressure. Through comprehensive analysis, we propose the prototypical electrical properties of NbO, for which a consensus has not yet been made. This study unequivocally demonstrates the utility of high temperatures in the thin film synthesis of refractory metal compounds.

\end{abstract}
\maketitle
\section{\label{sec:level1}Introduction} 

The interplay between electron correlations and strong spin-orbit coupling (SOC) has resulted in a variety of emergent quantum phenomena and topological phases \cite{witczak-krempaCorrelatedQuantumPhenomena2014}. Among prevailing material platforms, the 4\textit{d} and 5\textit{d} transition metal systems are ripe hosts for these effects, as SOC is a competing and comparable energy scale to kinetic energy. Some prominent examples in this correlated SOC regime include the 2D transition metal dichalcogenides \cite{manzeli2DTransitionMetal2017a}, Weyl semimetal pnictides \cite{armitageWeylDiracSemimetals2018a, yanTopologicalMaterialsWeyl2017}, and iridates \cite{rauSpinOrbitPhysicsGiving2016,witczak-krempaCorrelatedQuantumPhenomena2014}, some of which have hosted novel transport phenomena, such as extremely large nonsaturating magnetoresistance \cite{aliLargeNonsaturatingMagnetoresistance2014,shekharExtremelyLargeMagnetoresistance2015} and chiral anomaly induced negative longitudinal magnetoresistance \cite{huangObservationChiralAnomalyInducedNegative2015,zhangSignaturesAdlerBell2016}. However, these transport signatures can be obscured or complicated by extrinsic effects due to material imperfections or geometric configuration, such as weak localization \cite{luWeakAntilocalizationLocalization2015}, current jetting \cite{arnoldNegativeMagnetoresistanceWelldefined2016,yangCurrentJettingDistorted2019,lvExperimentalPerspectiveThreedimensional2021}, and ionic impurity induced scattering \cite{goswamiAxialAnomalyLongitudinal2015a}. Therefore, a thorough understanding of the mechanisms behind these interesting transport features and the synthesis conditions for reproducible, high-quality crystals, is essential.


NbO, one such transition metal compound, is a highly metallic 4\textit{d}$^{3}$ system and was among the first oxide superconductors discovered (\Tc~=~1.61~K) \cite{hulmSuperconductivityTiONbO1972,okazSpecificHeatMagnetization1975,pollardElectronicPropertiesNiobium1968}. NbO$_{1+x}$ forms in a unique vacancy-ordered rock-salt structure with 25\% vacancies on each sublattice \cite{burdettNiobiumOxideNbO1984}, and it exists over a very narrow homogeneity range (-0.02 $\leq x \leq$ 0.02) \cite{hulmSuperconductivityTiONbO1972,pollardElectronicPropertiesNiobium1968}, making it a controlled system for studying the influence of oxygen vacancy concentration and band filling on electrical transport properties. It has a complex band structure, with electrons and holes in several bands contributing to the transport properties \cite{wimmerEffectVacanciesElectronic1982,wahnsiedlerEnergyBandStructure1983,aokiFermiSurfaceNbO1990,efimenkoElectronicSignatureVacancy2017}. In bulk crystals, the transport properties were found to vary significantly with small changes in NbO$_{1+x}$ composition \cite{chandrashekharElectricalResistivityNbO1970,honigElectricalPropertiesNbO1973}, but there are many discrepancies between reports in terms of the magnitudes and trends, making interpretation difficult. These discrepancies can be attributed to impurities and the difficulty of precisely controlling and accurately determining the stoichiometry. 

In this work, we report the correlation between growth conditions and transport properties of NbO thin films in the oxygen partial pressure (\Po) and growth temperature (\Tg) phase space. At high growth temperatures (\Tg~$>$ 1000 \dC), accessible via laser heating, we observe a thermally-activated epitaxy window that yields sharp boundaries between NbO and its proximate phases, Nb and NbO$_{2}$. In this growth regime, we find superior crystal quality and transport properties, which are reproducible across a wide \Po~window. We propose the prototypical electrical properties, namely the temperature dependence of the Hall coefficient (\RH) and the superconducting transition temperature (\Tc), of NbO, for which a consensus has not yet been made.

\section{Methods}

Nb-O films were prepared on \AO~(0001) substrates using a molecular beam epitaxy system equipped with a CO$_2$ laser heating apparatus. \AO~(0001) was previously determined to be the most suitable substrate on which to grow NbO (111), despite the large lattice mismatch of $-7.7\%$ \cite{kimSuperconductingVacancyOrderedRockSalt2025}. Prior to growth, the substrates were annealed at $T$ = 1500~$^\circ$C for 10 minutes to obtain atomically flat terraces. Using a similar growth recipe to Ref. \cite{kimSuperconductingVacancyOrderedRockSalt2025}, a Nb flux of approximately 10 ng/cm$^{2}\cdot$s, as measured by a quartz crystal microbalance (QCM), was supplied by an electron beam evaporator. This flux resulted in an average NbO film thickness of $\approx$50 nm for a growth time of 1 hour. The flux was monitored during the growth using the QCM and the current supplied to the electron beam evaporator was adjusted to compensate for changes in Nb flux. Molecular oxygen was supplied using two variable leak valves in series with a baratron in between. The pressure in the intermediate volume was used as a tunable parameter. The corresponding chamber pressure using a nude ion gauge in the growth chamber ranged from 3 x 10$^{-8}$ to 1 x 10$^{-6}$ mbar for oxygen nozzle pressures of 0.05 to 0.40 mbar. 

X-ray diffraction (XRD) analysis was performed using a Rigaku Smartlab diffractometer with a Cu-$K\alpha1$ source. Reciprocal space mapping data was obtained using the Rigaku HyPix 3000 1D detector.

Electrical properties were measured using a Quantum Design Dynacool Physical Property Measurement System equipped with a dilution refrigerator option. The samples were wire bonded in a Van der Pauw configuration using aluminum wires. Resistivity measurements were obtained using the internal resistance bridge of the Dynacool system at a current of 100 $\mu$A for normal state transport measurements and 10 $\mu$A for measurements in the dilution refrigerator. 

\section{Results}
A common theme among 4\textit{d} and 5\textit{d} compounds containing refractory metals is that the high melting points, which result from strong metallic bonding \cite{martienssenSpringerHandbookCondensed2006b,campbellElementsMetallurgyEngineering2008b, savitskiiPhysicalMetallurgyRefractory2012}, create challenges when it comes to synthesis. Very high temperatures and significant time can be required in order for the metallic bonds to be broken and the elements to diffuse and react \cite{stormsRefractoryCarbides1967,tothTransitionMetalCarbides1971,piersonHandbookRefractoryCarbides1996}. Therefore, it can be challenging to realize ideal thermal conditions for the growth of superior-quality crystals and, given the additional restrictions in the thin film deposition process, their epitaxial growth. 

\begin{figure*}
\includegraphics[width=170 mm]{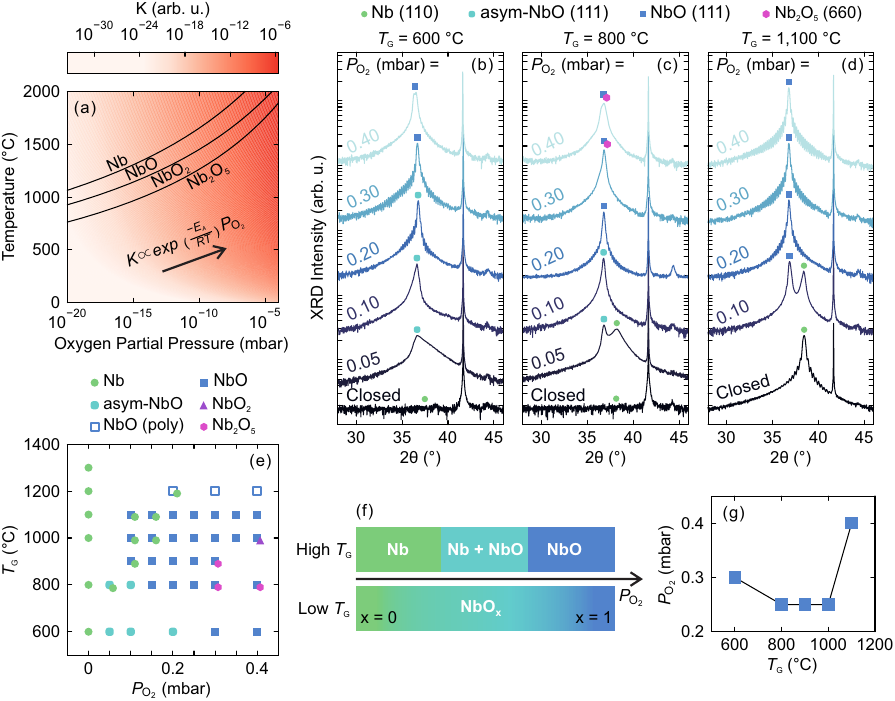}
\caption{\label{Fig1}Growth dynamics of the Nb-O system. (a) Kinetic and thermodynamic considerations for the formation of Nb oxides. The shaded contour plot represents the reaction rate ($K$) for the oxidation of Nb for various values of temperature and oxygen partial pressure. The black curves represent the Ellingham diagram for the formation of NbO, NbO$_2$, and Nb$_2$O$_5$ from the constituent elements. Thermodynamic data have been taken from Refs. \cite{chaseNISTJANAFThemochemicalTables1998, jacobThermodynamicPropertiesNiobium2010} using standard conditions. (b-d) XRD $2\theta$-$\omega$ scans of Nb-O films grown at \Tg~= 600, 800 and 1100 \dC~with varying \Po. (e) Growth phase diagram of Nb-O on \AO~(0001) in the~\Tg-\Po~parameter space. Symbols correspond to body-centered-cubic Nb (green circle), asymmetric NbO (cyan rounded square), polycrystalline NbO (open blue square), single-phase, single-crystal NbO (filled blue square),  rutile NbO$_2$ (purple triangle), and Nb$_2$O$_5$ (pink hexagon). (f) Schematic illustrating the effect of \Tg~on the amount of phase separation between Nb and NbO. (g) Optimal \Po~for forming the highest crystal quality NbO as a function of \Tg.}
\end{figure*}

Before describing the experimental details of our study, we consider two synthesis aspects. First, we consider equilibrium thermodynamics, which predict monotonically stronger reducing conditions as \Tg~increases and \Po~decreases \cite{shangEllinghamDiagramsBinary2024a}. To visualize this concept, we use the Ellingham diagram \cite{ellinghamTransactionsCommunications1944}, which relies on the relation between the Gibbs free energy change ($\Delta G$) and the equilibrium constant ($K$) under isobaric conditions,
\begin{equation} \label{eq1}
\Delta G = \Delta G_{0} + RT \textrm{ln} K = \Delta G_{0} - RT \textrm{ln} a_\mathrm{O_2}.
\end{equation}
Here, $\Delta G_{0}$ is the change in the standard Gibbs free energy, $R$ is the gas constant, $T$ is the absolute temperature, and $a_\mathrm{O_2}$ is the activity of oxygen. Second, we take into account kinetics, which is described by the Arrhenius equation. We have approximated the temperature-dependent reaction rate ($K$) for the oxidation of Nb as
\begin{equation} \label{eq2}
K \approx C^\prime \textrm{exp}(\frac{-E_{A}}{RT}) P_\mathrm{O_2}
\end{equation}
where $E_{A}$, the activation energy, is obtained from previous literature as 71 kJ/mol \cite{mclintockPressureDependenceLinear1963}, and the constant $C^\prime$ incorporates the Arrhenius factor and the activity of Nb, which is typically unity for solids \cite{shangEllinghamDiagramsBinary2024a}. These thermodynamic and kinetic mechanisms are illustrated in Fig. \ref{Fig1}(a). The black curves represent the Ellingham diagram for the oxidation of Nb, and the contour plot represents the reaction rate.

With these concepts in mind, we turn to the growth of the Nb-O thin film system. To control the oxidation state, we vary \Tg~and \Po~while keeping all other parameters fixed. In Figs. \ref{Fig1}(b-d) we show XRD $2\theta$-$\omega$ scans of Nb-O films grown at select \Tg~and varying \Po. At \Tg~= 600~\dC~[Fig. \ref{Fig1}(b)], amorphous Nb forms when the oxygen nozzle is closed. When the nozzle is open, bad-quality NbO, as evidenced by the asymmetric NbO (111) peak, forms at low \Po~(\Po~= 0.05 - 0.20 mbar). High-quality, symmetric NbO with well-defined Laue oscillations forms at intermediate \Po~(\Po~= 0.30 mbar). Further increasing \Po~to 0.40 mbar results in a slight splitting of the NbO (111) peak and a lack of Laue oscillations, indicating poor quality. At \Tg~= 800~\dC~[Fig. \ref{Fig1}(c)], amorphous Nb again forms at \Po~= ``Closed''. As \Po~is increased, the Nb-O system is progressively oxidized, transforming from bad-quality NbO mixed with Nb (\Po~= 0.05 mbar), to asymmetric NbO (\Po~= 0.10 mbar), to good-quality NbO (\Po~= 0.20 mbar), to NbO mixed with an impurity phase which is challenging to conclusively determine, but which is likely Nb$_2$O$_5$ (\Po~= 0.30 - 0.40 mbar) [Supplemental Material, Fig. S1]. At \Tg~= 1100~\dC~[Fig. \ref{Fig1}(d)], high-quality, crystalline Nb forms at \Po~= ``Closed'', and opening the nozzle to low \Po~(\Po~= 0.10 mbar) yields NbO mixed with Nb, with sharp separation between the peaks. Further increasing \Po~results in the formation of single-phase NbO with excellent crystal quality across a wide \Po~window (0.20 - 0.40 mbar). The crystal quality improves with increasing \Po, as evidenced by the increasing intensity and range of the Laue oscillations. 

\begin{figure*}
\includegraphics[width=170 mm]{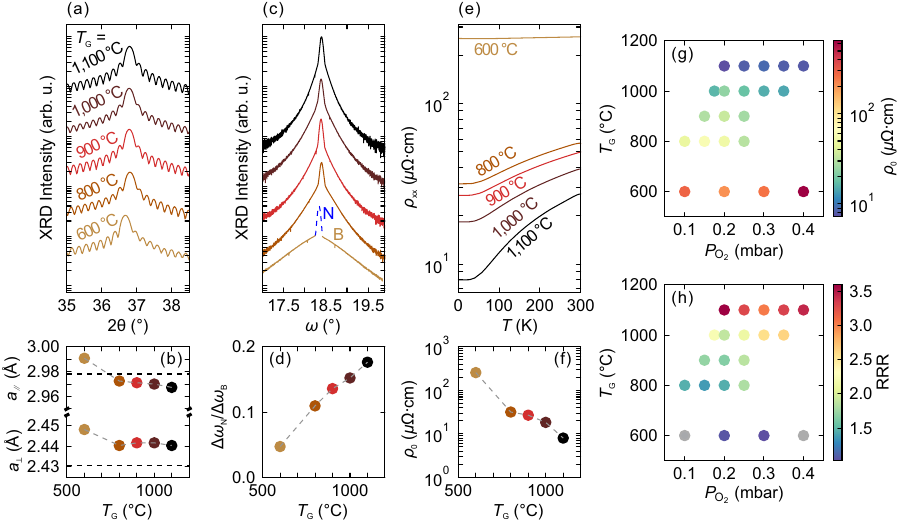}
\caption{\label{Fig2}Deterministic role of \Tg~on the structural and transport properties of NbO. (a) Narrow-range XRD $2\theta$-$\omega$ scans of NbO films grown at varying~\Tg~and the corresponding optimal \Po. (b) (Top) in-plane [11$\overline{2}$] lattice constant ($a_\sslash$) and (bottom) out-of-plane [111] lattice constant ($a_\perp$) of the same samples as a function of \Tg. Black dashed lines correspond to the respective bulk NbO values. (c) Rocking curve measurements of the NbO (111) peak and (d) ratio of the full width at half maximum (FWHM) of the narrow component (N) to the FWHM of the broad component (B) of the rocking curves. The curves were fitting using a Gaussian function for the narrow component and a squared Lorentzian function for the broad component. (e) Temperature-dependent longitudinal resistivity and (f) residual resistivity (\rr) at 1.7 K of the same samples. (g) \rr~and (h) residual resistivity ratio (RRR) of NbO samples in the \Tg-\Po~parameter space. Gray circles represent insulating samples.}
\end{figure*}

It is difficult to identify a comprehensive trend from these three sample sets, so we explore the growth of Nb-O in an extended range of the \Po-\Tg~parameter space [Fig. \ref{Fig1}(e)]. All \Po~= ``Closed'' samples are Nb metal whose crystalline state evolves from amorphous to (110)-oriented as \Tg~increases. Above \Tg~= 1100~\dC, NbO is polycrystalline, presumably due to entropy and the large lattice mismatch between NbO and \AO~(0001). The asymmetric NbO phase appearing in the low \Tg~and low \Po~conditions is tentatively considered to be NbO with Nb defects [Fig. \ref{Fig1} (f)], which will be discussed in detail later. Progressive oxidation occurs for all \Tg~series with increasing \Po~and, notably, higher oxidation states emerge only under intermediate \Tg~conditions. In a similar manner, a comparison of \Tg-varied samples at fixed \Po~(0.2, 0.3, and 0.4 mbar) suggests that the oxidation state increases and then decreases as a function of \Tg. This non-monotonic temperature dependence of oxidation can be qualitatively interpreted as a competition between the two trends discussed earlier. At low temperatures, NbO is stable up to high \Po, but it is rapidly oxidized with increasing temperature, in accord with the Arrhenius equation. At high temperatures, NbO becomes more reduced with increasing \Tg~and decreasing \Po, following Ellingham thermodynamics. The growth temperature triggers a shift in the growth mode from kinetics-driven (\Tg~$\leq$~900 \dC) to thermodynamics-driven (\Tg~$>$~900 \dC), suggesting thermal activation of the Nb-O system. The crossover temperature from kinetics- to thermodynamics-driven appears to be closely correlated with evolution of the Nb crystalline state as well as the occurrence of the asymmetric NbO phase. In accord with the non-monotonic \Tg-dependence of oxidation, the \Po~for forming the highest crystal quality NbO varies non-monotonically with \Tg, as shown in Fig. \ref{Fig1}g). The ``optimal'' samples were selected by evaluating the NbO (111) peak shape and Laue oscillations for all growth conditions [Supplemental Material, Fig. S2], and while this is not physically rigorous, it will be used as a guide for discussions below.

Given the deterministic role of \Tg~in the epitaxial Nb-O system, we investigate its effects on the structural and transport properties of NbO. Fig. \ref{Fig2}(a) shows narrow-range XRD $2\theta$-$\omega$ scans of NbO films grown at the optimal \Po~[see Fig. \ref{Fig1}(g)] for each \Tg~series. In Fig. \ref{Fig1}(b), we plot the in-plane ($a_\sslash$) and out-of-plane ($a_\perp$) lattice constants,  which are extracted from XRD analysis, as a function of \Tg. Both $a_\sslash$ and $a_\perp$ are larger than the bulk values for the \Tg~= 600 \dC~sample, but the lattice constants rapidly decrease as \Tg~increases, resulting in $a_\sslash$ being smaller and $a_\perp$ being larger than bulk, likely due to the influence of the substrate. Fig. \ref{Fig2}(c) shows the rocking curves of the NbO (111) peak of the same films. The rocking curves consist of a narrow (N) main peak and a broad (B) peak, the latter of which we attribute to a disordered region at the \AO-NbO interface. As \Tg~increases, the broad component becomes sharper, while the width of the narrow component remains similar. These changes can be captured by fitting the rocking curves using a Gaussian function for the narrow component and a squared Lorentzian function for the broad component [Supplemental Material, Fig. S3]. As shown in Fig. \ref{Fig2}(d), the ratio of the full width at half maximum of the narrow component to that of the broad component increases as \Tg~increases, indicating improvements in crystal quality at elevated growth temperatures. 

High \Tg~also helps to reproduce the low residual resistivity of NbO bulk crystals. Fig. \ref{Fig2}(e) shows the temperature-dependent longitudinal resistivity of the same samples. Aside from the \Tg~= 600 \dC~sample, the temperature dependence of the resistivity is similar to bulk samples, with an approximately linear $T$ dependence that flattens out due to residual disorder at low enough temperature \cite{honigElectricalPropertiesNbO1973,schulzBandStructureElectronic1992,pollardElectronicPropertiesNiobium1968}. As \Tg~increases, the temperature-dependent resistivity curve shifts downwards, which can be visualized by plotting the residual resistivity (\rr) at 1.7 K as a function of \Tg~[Fig. \ref{Fig2}(f)]. \rr~of the lowest resistivity sample in this series (\Tg, \Po) = (1100~\dC, 0.40 mbar) is 8 $\mu \Omega \cdot$cm, which is approaching the residual resistivity of 0.2 - 1.8 $\mu \Omega \cdot$cm in bulk NbO \cite{honigElectricalPropertiesNbO1973,schulzBandStructureElectronic1992,pollardElectronicPropertiesNiobium1968}. In Figs. \ref{Fig2}(g) and \ref{Fig2}(h), \rr~and the residual resistivity ratio (RRR) have been mapped out in the \Tg-\Po~parameter space. As \Tg~increases, \rr~decreases and RRR increases, and there is little dependence of these properties on \Po. 

\begin{figure*}
\includegraphics[width=170 mm]{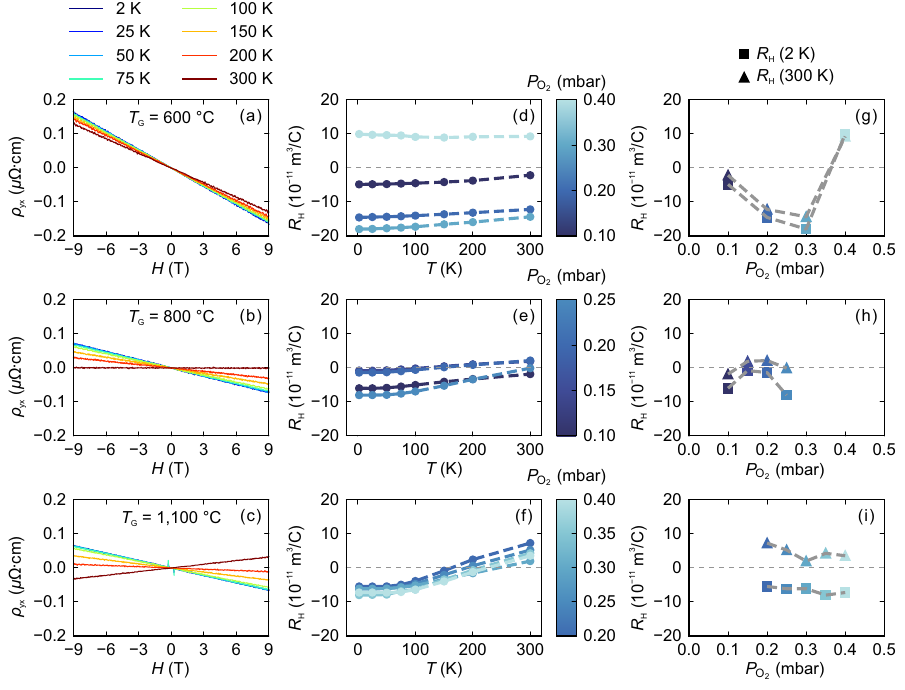}
\caption{\label{Fig3}Comparison of normal state transport properties of NbO films grown at~\Tg~=~600~\dC~(top row), 800 \dC~(middle row), and 1100~\dC~(bottom row). (a-c) Hall resistivity at different temperatures of NbO films grown at the optimal~\Po~for each \Tg. (d-f) Hall coefficient (\RH) versus temperature of samples grown at varying~\Po. (g-i) \RH~at 2 K (squares) and 300 K (triangles) of the same samples in (d-f). Dashed gray lines at \RH~= 0 serve as a guide to the eye.}
\end{figure*}

We now turn our attention to the transport properties of NbO which have had no consensus in the literature. The temperature dependence of \RH~has been reported for bulk NbO samples with conflicting results. In one report, \RH~changes sign from positive to negative as the measurement temperature increases \cite{schulzBandStructureElectronic1992}. In a different report, \RH~is negative and increases with rising temperature for oxygen-deficient and stoichiometric NbO, but it switches sign from negative to positive for oxygen-rich NbO \cite{honigElectricalPropertiesNbO1973}. There has also been disagreement about the magnitude of \Tc~and its dependence on the O/Nb ratio. In Ref. \cite{okazSpecificHeatMagnetization1975}, \Tc~reaches a maximum of 1.55 K at the stoichiometric composition and decreases at oxygen-deficient and oxygen-rich conditions. In Ref. \cite{pollardElectronicPropertiesNiobium1968}, \Tc~shows a qualitatively similar dependence on the O/Nb ratio, but it reaches a maximum of 1.45 K at slightly oxygen-rich conditions. On the other hand, Ref. \cite{hulmSuperconductivityTiONbO1972} reports a constant value of 1.38 K for stoichiometric and oxygen-rich conditions, and \Tc~rises sharply for oxygen-deficient conditions, which is attributed to the presence of Nb metal. Due to the effect of Nb impurities on the \Tc~of NbO, the magnitude of \Tc~is not a clear figure of merit in the NbO system.

In Figs. \ref{Fig3}(a-c), we plot the symmetrized Hall resistivity at different temperatures for the \Tg~= 600, 800, and 1100 \dC~samples, which are grown at the optimal \Po~for each \Tg. The unsymmetrized Hall resistivity for these samples can be found in Supplemental Material, Fig. S4. The slope of the Hall resistivity is negative at all temperatures for the \Tg~= 600 and 800~\dC~samples, whereas the slope changes from negative at low temperatures to positive at high temperatures for the \Tg~= 1100~\dC~sample. The temperature dependence of \RH~is illustrated in Figs. \ref{Fig3}(d-f), where we compare NbO samples grown at fixed \Tg~and varying \Po. At \Tg~= 600 \dC, \RH~is either positive at all temperatures or negative at all temperatures, depending on the \Po. At \Tg~= 800 \dC, the intermediate \Po~samples have a negative \RH~at low temperatures and positive \RH~at high temperatures, whereas the low- and high-\Po~samples exhibit a negative \RH~at all temperatures. At \Tg~= 1100 \dC, we observe the negative to positive temperature dependence of \RH, and the trend is consistent across all pressures. In Figs. \ref{Fig3}(g-i), we plot the \Po~dependence of \RH~at 2 K (squares) and 300 K (triangles). The \Tg~= 600~\dC~series shows a non-monotonic dependence of \RH~on \Po, first decreasing and then increasing as a function of \Po. The \Tg~= 800~\dC~series shows the opposite non-monotonic dependence, first increasing and then decreasing. In contrast, \RH~of the \Tg~= 1100~\dC~series at each temperature is nearly constant as a function of \Po. We note that in all cases, \RH~is similar in magnitude to bulk reports, but its dependence on \Po~follows the trends observed in Ref. \cite{honigElectricalPropertiesNbO1973}.

\begin{figure}[t]
\includegraphics[width=85 mm]{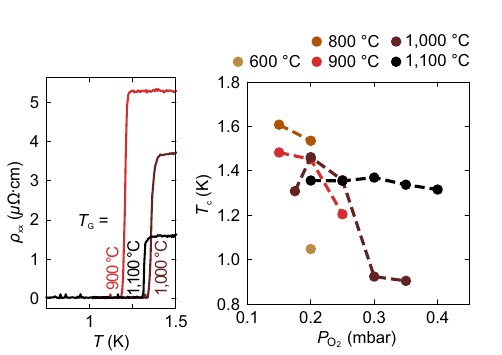}
\caption{\label{Fig4} Superconducting properties of NbO. (a) Temperature-dependent longitudinal resistivity of select samples grown at \Tg~= 900, 1000, and 1100 \dC~and the corresponding optimal \Po. (b) Superconducting transition temperature (\Tc) of NbO films as a function of~\Po~for samples grown at varying \Tg.}
\end{figure}

Fig. \ref{Fig4}(a) shows the temperature-dependent resistivity of NbO samples grown at \Tg~= 900, 1000, and 1100 \dC~and the optimal \Po. The transitions are all sharp, indicating good sample homogeneity. However, \Tc, defined as the temperature at half resistance, does not show a clear dependence on \Tg, so to gain insight we plot \Tc~as a function of \Po~for the different \Tg~series [Fig. \ref{Fig4}(b)]. For the \Tg~= 800 and 900~\dC~samples, \Tc~decreases slightly with increasing \Po. At \Tg~= 1000~\dC, \Tc~increases initially and then decreases rapidly before flattening out above \Po~= 0.3 mbar. In contrast, there is little change in \Tc~for the \Tg~= 1100~\dC~samples as a function of \Po.

\section{Discussion}

Our results establish that \RH~and \Tc~of NbO vary dramatically with growth parameters, and the beneficial effects of \Tg~on the quantifiable figures of merit indicate that thermal activation is indispensable to realizing the desirable properties of NbO films. Although 600$\sim$1000~\dC~is a conventional temperature range for most thin film deposition techniques, the ideal growth temperature for the epitaxial Nb-O system exceeds this window due to the refractory nature of Nb. 
To further this discussion, we consider a simple model of the Nb-O system in which the oxygen content varies, resulting in non-stoichiometric NbO$_{1+x}$ (-0.02 $\leq x \leq$ 0.02). When the amount of deviation is above that which the NbO phase can tolerate, the coexistence of NbO and a secondary phase occurs. These secondary phases are Nb on the oxygen-deficient side and NbO$_2$ on the oxygen-rich side. The manner in which non-stoichiometry and phase coexistence influence the electrical properties are discussed below. 

The superconductivity of NbO is likely of conventional origin \cite{hulmSuperconductivityTiONbO1972}, so \Tc~is positively correlated with the Debye temperature ($\Theta_\mathrm{D}$), the density of states at the Fermi level ($N_\mathrm{0}$), and the electron-phonon interaction ($V_\mathrm{ph}$) \cite{bardeenTheorySuperconductivity1957}. Within the NbO region, we expect an electron doping effect (Fermi level rise) as oxygen content decreases. DFT calculations show that $N_\mathrm{0}$ increases as the Fermi level rises \cite{wimmerEffectVacanciesElectronic1982,efimenkoElectronicSignatureVacancy2017}, which accordingly should influence \Tc. While conventional superconductors are generally tolerant to non-magnetic crystalline disorder, structural non-uniformity can suppress \Tc~and/or widen the superconducting transition [see Supplemental Material, Fig. S5 for transitions]. Combining the effects of $N_\mathrm{0}$ and crystallinity, we expect \Tc~to be maximized at a slightly sub-stoichiometric condition. Further deviating from the stoichiometric composition, in the Nb-rich region, \Tc~has been reported to rapidly increase due to the inclusion of Nb, which has a much higher \Tc~(9.25 K) than NbO \cite{finnemoreSuperconductingPropertiesHighPurity1966}. In the NbO$_2$-containing region, \Tc~is expected to be moderately suppressed as NbO$_2$ is not superconducting, resulting in a non-uniform superconducting landscape where pairing is suppressed via proximity to a normal material.

Turning to bulk transport properties, \RH~is an algebraic sum of the individual contributions of each carrier, which are of different sign \cite{wahnsiedlerEnergyBandStructure1983}. As NbO contains many electron and hole pockets near the Fermi level, shifting the Fermi level changes the relative filling of the bands. According to the band structure, NbO becomes hole-dominated as the Fermi level is decreased (oxygen content is increased) and electron-dominated as the Fermi level is increased (oxygen content is decreased). Therefore, in the NbO$_{1+x}$ region, \RH~should become more positive (negative) in oxygen-rich (deficient) conditions. As for the defect phases, \RH~is reported to be positive for both Nb \cite{berlincourtHallEffectResistivity1959,coxTemperatureDependenceHall1973}, and NbO$_2$ \cite{belangerElectronTransportSingle1974a,sakaiElectricalPropertiesSemiconducting1985}. 

These trends largely agree with our observations in the \Tg~= 1000 \dC~sample series. In the NbO$_{1+x}$ region, \Tc~is maximized at a pressure lower than the structurally optimal \Po, while it decreases on either side of this pressure. When \Po~is further decreased towards the Nb-rich region, \Tc~sharply increases [Supplemental Material, Fig. S6]. \RH~follows a similar trend as a function of \Po~[Supplemental Material, Fig. S7], which is likely due to the competing dominance of different carrier types in the various phase regions. 

In the low \Tg~samples, these trends are complicated by the lack of sharp phase boundaries between Nb and NbO [Fig. \ref{Fig1}(f)]. In the Nb-O system, we are able to detect Nb via electrical but not structural measurements [Supplemental Material, Fig. S6]. We can infer that this non-detectable Nb is more pervasive in the low-\Tg~films, given the blurring of the phase boundaries and the amorphous nature of Nb on \AO~(0001) at these temperatures. We clearly resolve that \Tc~increases as we reduce \Po, but it is difficult to distinguish the effects of Nb impurities and the NbO Fermi level change. \RH~does not show a clear dependence on \Tg~or \Po, and interpretation is difficult due to the extrinsic effects of poor crystallinity and phase coexistence. For the sake of completeness we have also measured \Tc~and \RH~of amorphous and crystalline Nb films [Supplemental Material, Fig. S8].

We finally discuss the role of high temperature synthesis in NbO, and, more generally, in reduced refractory oxides. First, as a basic means of improving adatom mobility and sample homogeneity, high \Tg~($>$ 1000 \dC) enables superior structural and electrical properties (evidenced by \rr~and RRR). Second, it facilitates the reproducible synthesis of NbO within the thermally-activated epitaxy regime where the magnitude of \Tc, and the sign and magnitude of \RH~at different temperatures, are consistent for a finite range of \Po-variation. This thermally-activated epitaxy closely resembles reported self-limited growth mechanisms that yield optimal material properties within a wide growth window. Examples include the adsorption-controlled growth of GaAs \cite{arthurInteractionGaAs21968} and SrTiO$_3$ \cite{jalanMolecularBeamEpitaxy2009a}, and the diffusion-controlled growth of the Ti-O system \cite{kimHighTemperatureDiffusionEnabled2025a}.

\section{Conclusion}

We observed competing growth mechanisms in the synthesis of the Nb-O system, driven by kinetics at low temperatures and thermodynamics at high temperatures. Ultra-high growth temperature was a key to realizing excellent crystal quality, superior transport properties, and reproducible synthesis. We have demonstrated improved quality of NbO at ultra-high temperatures and a growth window yielding consistent physical properties. As such, we propose the following prototypical properties of NbO: (1) \RH~that is negative at low temperatures and positive at high temperatures, and (2) \Tc~of 1.32-1.37 K. Our study highlights the importance of thermal activation in the thin film synthesis of refractory metal compounds across large thermodynamic windows.

\section*{Acknowledgments}
We thank Bharat Jalan for his valuable comments. We acknowledge funding provided by the Gordon and Betty Moore Foundation’s EPiQS Initiative (Grant number GBMF10638), the Institute for Quantum Information and Matter, a NSF Physics Frontiers Center (NSF Grant PHY-2317110), and the AWS Center for Quantum Computing through a sponsored research initiative. This research was carried out at the Jet Propulsion Laboratory and the California Institute of Technology under a contract with the National Aeronautics and Space Administration and funded through the President’s and Director’s Research \& Development Fund Program.

\bibliography{ref.bib}

\clearpage

\input{SI_arxiv} 

\end{document}

%% file: SI_arxiv.tex
\title{Supplemental Material\\ 
``Thermally-Activated Epitaxy of NbO"}
\author{Sandra~Glotzer}
\affiliation{Department of Applied Physics and Materials Science, California Institute of Technology, Pasadena, California 91125, USA.}
\affiliation{Institute for Quantum Information and Matter, California Institute of Technology, Pasadena, California 91125, USA.}

\author{Jeong~Rae~Kim}
\affiliation{Department of Applied Physics and Materials Science, California Institute of Technology, Pasadena, California 91125, USA.}
\affiliation{Institute for Quantum Information and Matter, California Institute of Technology, Pasadena, California 91125, USA.}

\author{Joseph~Falson}
\email{falson@caltech.edu}
\affiliation{Department of Applied Physics and Materials Science, California Institute of Technology, Pasadena, California 91125, USA.}
\affiliation{Institute for Quantum Information and Matter, California Institute of Technology, Pasadena, California 91125, USA.}

\maketitle


\renewcommand{\thefigure}{S\arabic{figure}}
\setcounter{figure}{0}

\begin{figure*}[t]
  \centering
   \includegraphics[width=85mm]{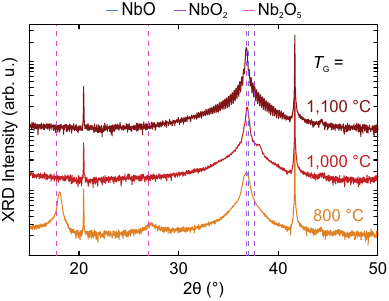}
   \caption{Identification of impurity phases in the Nb-O system. Wide-range XRD $2\theta$-$\omega$ scans of Nb-O films grown at \Po~= 0.4 mbar and varying \Tg. Dashed lines correspond to the bulk peak positions for NbO (blue), rutile NbO$_2$ (purple), and H-Nb$_2$O$_5$ (pink). As there are numerous polymorphs and crystal structures of Nb$_2$O$_5$, it is difficult to unambiguously assign a phase.}
   \label{FigS1}
\end{figure*}

\begin{figure*}[t]
  \centering
   \includegraphics[width=170mm]{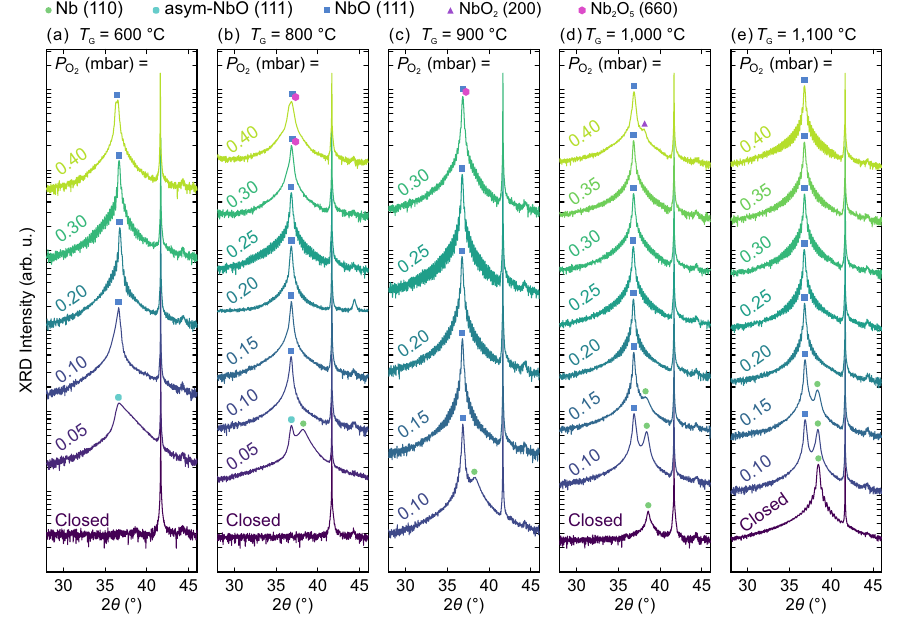}
   \caption{Narrow-range XRD $2\theta$-$\omega$ scans of Nb-O films grown at varying \Po~and (a) \Tg~=~600~\dC, (b) \Tg~=~800~\dC, (c) \Tg~=~900~\dC, (d) \Tg~=~1000~\dC, and (e) \Tg~=~1100~\dC. Symbols correspond to body-centered-cubic Nb (green circle), asymmetric NbO (cyan rounded square),  single-phase, single-crystal NbO (filled blue square),  rutile NbO$_2$ (purple triangle), and Nb$_2$O$_5$ (pink hexagon).}
   \label{FigS2}
\end{figure*}

\begin{figure*}[t]
  \centering
   \includegraphics[width=170mm]{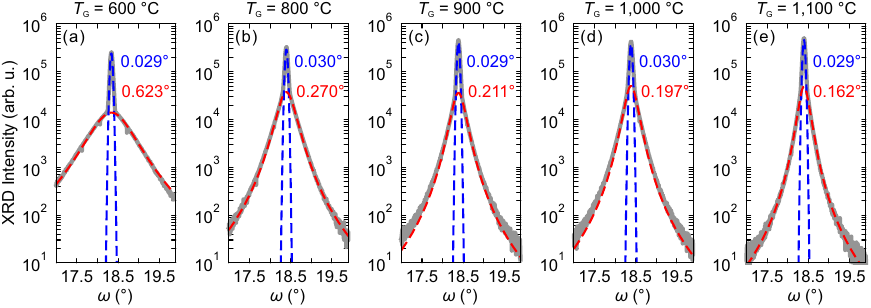}
   \caption{Fits of the  NbO (111) rocking curves for the samples shown in Fig. 2. Samples were grown at the optimal \Po~for each \Tg~and (a) \Tg~=~600~\dC, (b) \Tg~=~800~\dC, (c) \Tg~=~900~\dC, (d) \Tg~=~1000~\dC, and (e) \Tg~=~1100~\dC. The curves were fitted using a Gaussian function (blue) and a squared Lorentzian function (red). The full width at half maximum values of the Gaussian and squared Lorentzian components are displayed in blue and red text.}
   \label{FigS3}
\end{figure*}

\begin{figure*}[t]
  \centering
   \includegraphics[width=170mm]{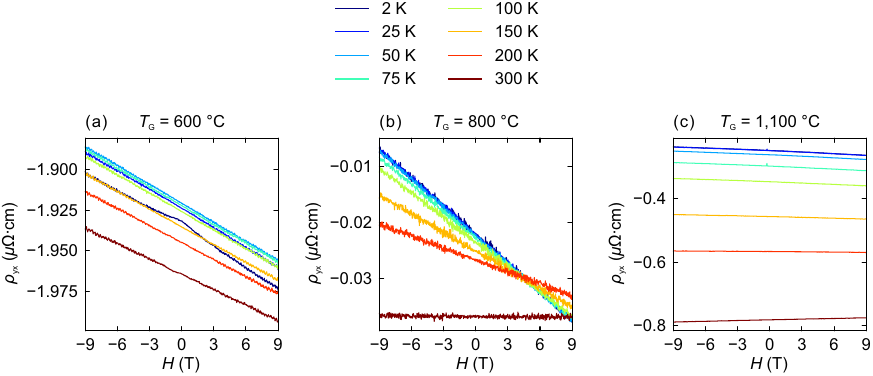}
   \caption{Unsymmetrized Hall resistivity for the NbO samples shown in Fig. 3. Samples were grown at the optimal~\Po~for each \Tg~and (a) \Tg~=~600~\dC, (b) \Tg~=~800~\dC, and (c) \Tg~=~1100~\dC~.}
   \label{FigS4}
\end{figure*}

\begin{figure*}[t]
  \centering
   \includegraphics[width=170mm]{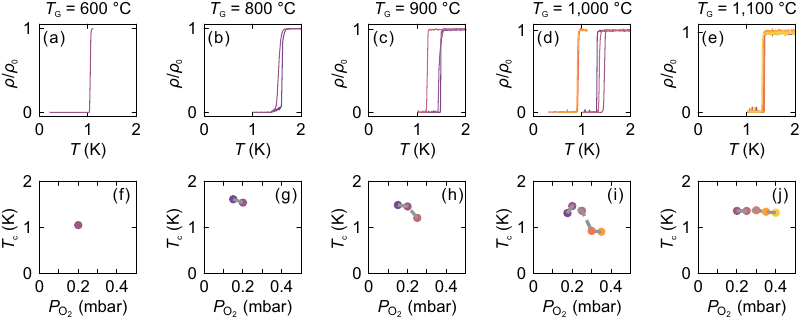}
   \caption{Superconducting transitions of NbO films grown at~\Tg~=~600~\dC~(first column), \Tg~=~800~\dC~(second column), \Tg~=~900~\dC~(third column), \Tg~=~1000~\dC~(fourth column),  and ~\Tg~=~1100~\dC~(fifth column). (a-e) Temperature-dependent longitudinal resistivity at various \Po~and (f-j) \Tc~as a function of \Po.}
   \label{FigS5}
\end{figure*}

\begin{figure*}[t]
  \centering
   \includegraphics[width=85mm]{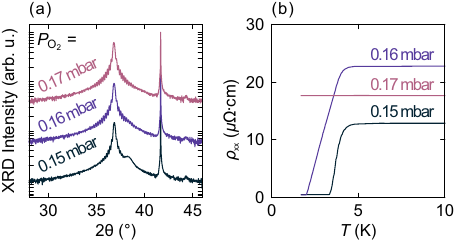}
   \caption{Narrow-range \Po-dependent growths of Nb-O films grown at \Tg~= 1000 \dC. (a) XRD $2\theta$-$\omega$ scans and (b) temperature-dependent longitudinal resistivity.}
   \label{FigS6}
\end{figure*}

\begin{figure*}[t]
  \centering
   \includegraphics[width=170mm]{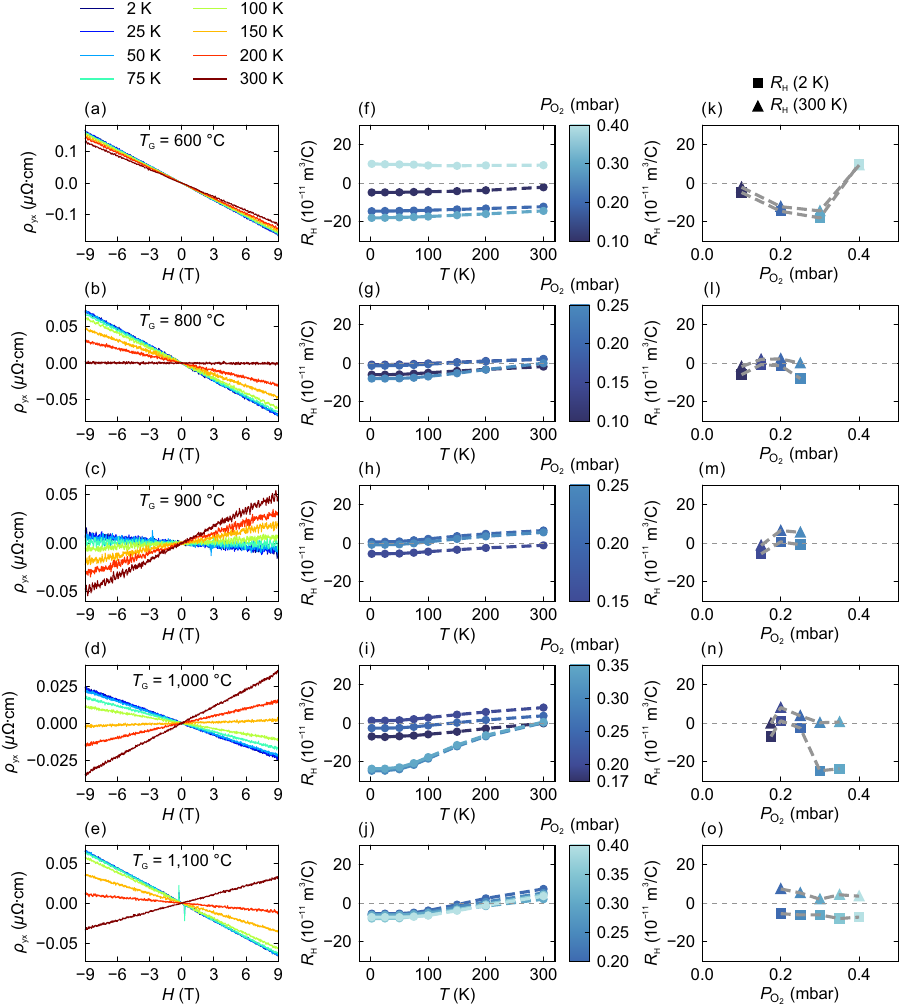}
   \caption{Comparison of transport properties of NbO films grown at~\Tg~=~600~\dC~(first row), \Tg~=~800~\dC~(second row), \Tg~=~900~\dC~(third row), \Tg~=~1000~\dC~(fourth row),  and ~\Tg~=~1100~\dC~(fifth row). (a-e) Hall resistivity at different temperatures of NbO films grown at the optimal~\Po~for each \Tg. (f-j) Hall coefficient (\RH) versus temperature of samples grown at varying~\Po. (k-o) \RH~at 2 K (squares) and 300 K (triangles) of the same samples. Dashed gray lines at \RH~= 0 serve as a guide to the eye.}
   \label{FigS7}
\end{figure*}

\begin{figure*}[t]
  \centering
   \includegraphics[width=85mm]{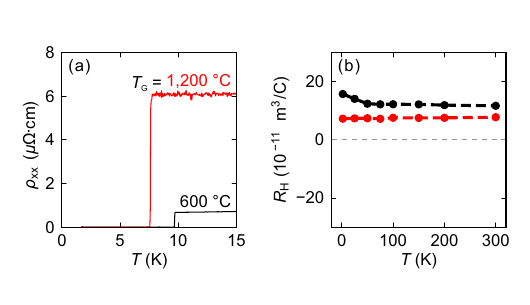}
   \caption{Transport measurements of an amorphous Nb film grown at (\Tg, \Po) = (600~\dC, ``Closed'') (black) and a crystalline Nb film grown at (\Tg, \Po) = (1200~\dC, ``Closed'') (red). a) Temperature-dependent longitudinal resistivity. (b) \RH~versus temperature. Dashed gray line at $R_H$ = 0 serves as a guide to the eye.}.
   \label{FigS8}
\end{figure*}